\documentclass{aastex}
\usepackage{emulateapj5,danonecolfloat}

\newcommand{\polar} {{\scshape Polar }}
\newcommand{\polars} {{\scshape Polar's }}

\newcommand{\lan}{\langle}
\newcommand{\ran}{\rangle}
\newcommand{\bi}{\begin{itemize}}
\newcommand{\ei}{\end{itemize}}

\def\GHz{{\rm GHz}}

\def\expec#1{\langle#1\rangle}

\def\ie{{\frenchspacing\it i.e.}}
\def\eg{{\frenchspacing\it e.g.}}

\def\beq#1{\begin{equation}\label{#1}}
\def\eeq{\end{equation}}
\def\beqa#1{\begin{eqnarray}\label{#1}}
\def\eeqa{\end{eqnarray}}
\def\eq#1{equation~(\ref{#1})}

\def\x{{\bf x}}

\def\rh{\widehat{\bf r}}

\def\C{{\bf C}}

\def\M{{\bf M}}

\def\R{{\bf R}}

\def\l{\ell}

\def\ith{i^{th}}

\shorttitle{\polar 2000 Results}

\begin{document}



\twocolumn[

\title{A Limit on the Large Angular Scale Polarization of the Cosmic Microwave Background}


\author{Brian G. Keating\altaffilmark{1}, Christopher W. O'Dell,
Angelica de Oliveira-Costa\altaffilmark{2}, Slade Klawikowski,
Nate Stebor\altaffilmark{3}, Lucio Piccirillo\altaffilmark{4},
Max Tegmark\altaffilmark{2}, and Peter T. Timbie}
\affil{Department of Physics, University of Wisconsin -- Madison,
Madison, WI 53706}


\begin{abstract}
We present an upper limit on the polarization of the Cosmic
Microwave Background at $7\arcdeg$ angular scales in the
frequency band between 26 and 36 \GHz, produced by the \polar
experiment. The campaign produced a map of linear polarization
over the R.A. range $112\arcdeg - 275\arcdeg$ at declination
$43\arcdeg$. The model-independent upper limit on the E-mode
polarization component of the CMB at angular scales $\ell = 2 -
20$ is $10 \mu$K (95\% confidence). The corresponding limit for
the B-mode is also $10 \mu$K. Constraining the B-mode power to be
zero, the $95\%$ confidence limit on E-mode power alone is
$8\mu$K.
\end{abstract}


]

\altaffiltext{1}{Current Address: Department of Physics,
California Institute of Technology, Pasadena, CA 91125;
bgk@astro.caltech.edu.} \altaffiltext{2}{Department of Physics,
University of Pennsylvania, Philadelphia, PA 19104}
\altaffiltext{3}{Department of Physics, University of California
at Santa Barbara, Santa Barbara, CA 93106}
\altaffiltext{4}{Department of Physics and Astronomy, University
of Wales - Cardiff, Wales, UK CF24 3YB}

\section{Introduction}
The Cosmic Microwave Background (CMB) is completely specified by
three characteristics: its spectrum, the spatial distribution of
its total intensity, and the spatial distribution of its
polarization. In the past decade, measurements of its spectrum
and temperature anisotropy have ushered in an era of precise
determinations of numerous cosmological parameters. CMB
polarization has the potential to provide powerful cross-checks as
well as to improve the accuracy with which cosmological parameters
can be measured, most strikingly those related to gravitational
waves and the ionization history of the Universe. Both of these
phenomena imprint a signature on the polarization of the CMB at
the largest angular scales $\ell \la 30$. We have built a
polarimeter called Polarization Observations of Large Angular
Regions ({\scshape Polar}) which is optimized to study CMB
polarization at these scales.

The polarization of the CMB is a unique probe of radiative
transport in the pre-galactic plasma. Thomson scattering of
anisotropic CMB radiation by free electrons produces CMB
polarization, which therefore carries information about the
Universe during the periods when it was ionized. Information
about the early ($z\la 1000$) ionized epoch is encoded in
degree-scale polarized fluctuations, whereas information about
reionization is imprinted on scales of tens of degrees,
corresponding to the horizon scale at the time. Therefore, CMB
polarization provides the crucial evolutionary link from the
recombination epoch ($z \sim 1000$) to the period of large scale
structure formation ($z \sim 6$). In particular, reionization
should produce a new polarized peak near $\ell\la 20$, the peak
location depending on the redshift $z_i$ at which the first
luminous objects heated and reionized the Universe
\citep{zal98,kea98}. Despite its fundamental importance, $z_i$ is
still quite poorly constrained observationally: the lack of an
observed Lyman-$\alpha$ trough (due to neutral hydrogen) in the
spectra of distant quasars \citep{gun65} places a lower limit
$z_i\ga 6$, whereas the lack of observed suppression of the
unpolarized acoustic peaks gives an upper limit $z_i\la 20$
\citep{wan01}. The purpose of this {\it Letter} is to present the
strongest limits to date on CMB polarization on the angular scales
$\ell\la 30$ relevant to these polarized peaks.

\section{Instrument}
\label{sec:instrument} \polars design builds on techniques
developed in previous searches for CMB polarization
\citep{nan79,lub81,wol93} and is driven by the magnitude and
angular scale of the anticipated CMB signal, and rejection of
potential systematic effects. The polarimeter is a
superheterodyne correlation radiometer which detects two
orthogonal linear polarization states in three radiofrequency
(RF) bands in the $K_\textrm{a}$-band between 26-36 GHz. The two
polarization states $i \in \{x,y\}$, $E^{RF}_{i}(t,\nu,\phi_{i})=
E_{i}\cos ( 2\pi\nu t+\phi_{i})$ enter a single-mode corrugated
feedhorn and are separated by an orthomode transducer (OMT). The
OMT's high polarization isolation ($>30$ dB) ensures low
cross-polarization. $E^{RF}_i$ are amplified by separate HEMT
amplifiers \citep{pos} cooled to 25 K by a mechanical cryocooler.
Downconversion from the RF band to an intermediate frequency (IF)
band (2-12 GHz) is performed by Schottky diode mixers, driven by
a Gunn diode local oscillator (LO) at 38 GHz. In the IF band the
two polarization states are amplified producing $E^{IF}_{i}
\propto E^{RF}_{i}$, and filtered into three separate IF bands,
denoted J1, J2, and J3. The IF bands translate into RF bands: J1
(32-36 GHz), J2 (29-32 GHz), and J3 (26-29 GHz). Prior to
filtering, two diode detectors measure the total power of each
polarization state, which monitors atmospheric opacity.
$E^{IF}_{x}$ and $E^{IF}_{y}$ are correlated by three Schottky
Diode analog multiplier circuits. The phase of the LO is switched
between $0$ and $\pi$ at 1 KHz prior to mixing the $E^{RF}_y$
waveform. The voltage produced by the correlators at this stage
thus switches between $\kappa E^{RF}_x E^{RF}_y $ and $-\kappa
E^{RF}_x E^{RF}_y $ at 1 KHz, where $\kappa$ is the
intensity-to-voltage conversion factor. Phase-sensitive detection
of this modulated signal reduces the effects of low frequency
drifts in the HEMTs and subsequent components to negligible
levels. After low-pass filtering, we record an audio-band signal
with a DC component $I_{DC} = \kappa \lan E^{RF}_x E^{RF}_y \ran
$, where the brackets denote a time-average, and AC
components proportional to the thermal noise from the radiometer,
atmosphere, and celestial signals. These signals are referred to
as the ``science channels''. A second lock-in amplifier for each
correlator is referenced to the same 1 KHz waveform but delayed
in phase by $\pi/2$ with respect to the phase switch. These
signals, hereafter referred to as ``quadrature phase channels''
(QPC), contain only the noise terms of the RF band and no optical
or celestial signals. The QPC are powerful probes of systematic
effects produced solely by the radiometer and post-detection
stages. The output from the QPC is proportional to the noise
equivalent temperature (NET) of the instrument. For a correlation
radiometer this is $\rm{NET} =
\sqrt{2/\Delta\nu}(T_{\rm{Rx}}+T_{\rm{Ant}})[\rm{K-s^{1/2}]}$,
where $\Delta\nu$ is the bandwidth of the radiometer,
$T_{\rm{Rx}}$ is the receiver noise temperature, and
$T_{\rm{Ant}}$ is the antenna temperature of observed optical
sources, including diffuse sources such as the atmosphere and the
CMB itself. A schematic outline of \polar is presented in
\citet{kea98}, and \citet{kea00} presents the complete design. The
specifications of \polar are summarized in Table 1.

\begin{deluxetable}{cccccc}
\small
 \label{t:polarspex}
 \tablewidth{270pt}
 \tablecaption{\polar Instrument Specifications}
 \tablehead{
& \colhead{$\nu_c$\tablenotemark{b}}&
\colhead{$\Delta\nu$\tablenotemark{c}}& \colhead{FWHM
\tablenotemark{d}}&
\colhead{$\overline{T_{\rm{pol}}}$\tablenotemark{e}}&
\colhead{$S_{\rm{sky}}$\tablenotemark{f}}\\
\colhead{Channel\tablenotemark{a}} & \colhead{[GHz]} &
\colhead{[GHz]} &
\colhead{[$\arcdeg$]}&\colhead{[$\mu$K]}&\colhead{[$\rm{mK}-s^{1/2}$]}\\
}
 \startdata
\small{TP-E/H} & 31.9 & \small{7.8/8.0}&  7.0 &  \nodata & \small{14.0/20.0}  \\
J3 & 27.5 &  0.8 & 7.7 &  84(28) & 1.9 \\
J2 & 30.5 & 2.0 &  7.4 &  72(14) & 1.0 \\
J1 & 34.0 & 1.9 &  7.1 &  33(11) & 1.0  \\
\enddata
\tablenotetext{a}{\footnotesize{TP-E and TP-H measure the total
power in the E and H polarization planes of the Horn/OMT assembly
prior to correlation.}} \tablenotetext{b}{\footnotesize{Channel
Band Centroid}} \tablenotetext{c}{\footnotesize{Channel
Bandwidth.}} \tablenotetext{d}{\footnotesize{E and H-plane
Beamwidths are equal to within $5\%$. Measured feed/OMT
cross-polarization is $<-40$dB for all channels.}}
\tablenotetext{e}{\footnotesize{Mean Polarized Offset
$\overline{T_{\textrm{pol}}}=\sqrt{\overline{Q}^2+\overline{U}^2}$,
where $\overline{Q}$ and $\overline{U}$ are the Stokes parameter
offsets. Numbers in parentheses denote the corresponding values
for the QPC.}} \tablenotetext{f}{\footnotesize{Measured channel
NET for a typical clear day with $K_{\rm{a}}$-band zenith sky
temperature $T_{Atm} \simeq 15$ K. NET measured at Stokes
modulation frequency $0.065$ Hz.}}
\end{deluxetable}

Absolute calibration of the polarimeter is accomplished by
placing a 0.003" polypropylene sheet in the near-field of the
antenna, tilted at an angle of $45\arcdeg$ relative to the optical
axis. The dielectric sheet introduces a polarized signal into the
antenna due to the difference in reflection between the parallel
and perpendicular electric fields. We calibrate the polarimeter by
rotating it around the optical axis. Near-field, absolute
calibration of previous CMB polarimeters have used wire-grids to
reflect the fields parallel to the wire axes \citep{lub81}. The
dielectric sheet material and thickness are chosen to produce an
absolute polarized antenna temperature of $\sim 500$ mK versus
$~200$ K for wire-grid calibrators. This provides a calibration
reference near the NET of the radiometer (accurate to ~$10\%$),
which reduces the dynamic range requirements on the IF and
post-detection stages. A $\chi^2$ minimization fit to a
radiometer model is computed during calibration, which determines
not only the radiometer flux scale, but also inter-channel
cross-talk and gain imbalance between the two polarization states
\citep{kea00}.

\section{Observations}
The \polar campaign collected a total of 746 hours of data between
2000 March 11 and 2000 May 29. Both Stokes parameters $Q$ and $U$
were observed through the same atmospheric column by a single
corrugated scalar feedhorn cooled to $\sim 40$K. The feed
produces a nearly frequency independent beam with a FWHM $\simeq
7.0 \arcdeg$. The beam observes the zenith from Madison,
Wisconsin, which corresponds to a declination of $\delta =
43\arcdeg 01\arcmin 48 \arcsec$. Contamination by terrestrial
radiation is minimized by our choice of vertical drift scans and
is further mitigated by two levels of ground shields (one
co-rotating with the polarimeter, and the other fixed relative to
the ground). Thus, radiation from the ground must diffract over
two ground screens before reaching the antenna.

The Stokes parameters completely describe the linear polarization
state of radiation produced by Thomson scattering
\citep{cha60,hu97} and are modulated by rotation of the entire
instrument around its optical axis. The radiometer rotates at $f =
0.033$Hz. The Stokes parameters are modulated at $2f$ whereas most
potential systematic signals are modulated at $f$ or DC, and are
easily discriminated from $Q$ and $U$. Measurement of both $Q$
and $U$ is essential for model independent recovery of the
`Electric' (or `Gradient') and `Magnetic' (or `Curl') components
of the polarization power spectrum \citep{zal97,kks97}.

\section{Data Reduction and Maps}
\label{section:data} The data were sampled at 20 Hz and converted
from digital units to Stokes $Q$ and $U$ for each 7.5 minute
file. Each file contains $\sim 14$ rotations of the polarimeter
during which the beam moves $\sim 1.9\arcdeg$. For each file, the
data are sorted into discrete angular bins and synchronously
demodulated with respect to trigonometric functions of the
rotation angle. In our coordinate system, defined with respect to
the local geographic coordinates, we perform a $\chi^2$
minimization of the angular binned data fit to the function
\beq{e:signal} I(\theta_t) = I_o + C\cos\theta_t + S\sin \theta_t
+ Q\cos 2\theta_t + U\sin 2\theta_t, \eeq where $\theta_t = 2\pi
f t$. In addition to the Stokes parameters $Q$ and $U$, the terms
$C$ and $S$ (which are synchronous with the rotation at frequency
$f$) are monitored to determine our sensitivity to
rotation-synchronous systematic effects, and to monitor
atmospheric fluctuations.

Three levels of diagnostics were used to detect and remove
contaminated data: 1) housekeeping and weather (dewpoint, cloud
cover) cuts; 2) time ordered data (TOD) cuts; 3) rotation ordered
data (ROD) cuts after constructing the Stokes parameters. Data
were collected while the sun (moon) was less than $30\arcdeg$
($60\arcdeg$) above the horizon. Approximately $2\%$ of the total
data collected were removed due to work on the receiver or
closure of the observatory dome. $1.5\%$ of the data were removed
due to completely overcast cloud cover.  Although atmospheric
emission is not expected to have significant linear
polarization at 30 GHz \citep{kea98}, the
correlators are sensitive to changes in the total antenna
temperature. This coupling is caused by imperfect isolation
between the E and H planes of the OMT and results in a weak
dependence of the correlators on their total bias power
\citep{car01}. The bias power is proportional to the system
temperature, including atmospheric emission. Since the
atmospheric fluctuation spectrum at 30 GHz falls approximately as
$1/f^\alpha$ ($\alpha \geq 1$), due to fluctuations in the amount
of precipitable water vapor, there is residual power at the
Stokes parameter modulation frequency $0.065$ Hz. This effect is
correlated between all frequency channels, and produces a slow
drift in the demodulated Stokes parameters, as well as in the
rotation synchronous terms $S$ and $C$ in equation
\ref{e:signal}. We require that the rotation synchronous terms
$S$ and $C$ be less than $2\times$NET. This cut (applied
after the TOD has been assembled into the ROD) removes almost all
weather-related contamination of the science channels ($35\%$ of
the total data). Since the basis functions for $S$ and $C$ are
orthogonal to those of $Q$ and $U$, we are assured that only data
collected during poor weather, and not polarized celestial
signals, are flagged. Further data files ($2.6\%$) are removed if
any samples in the file deviate more than $5\sigma$ from the
TOD-mean, or $3\sigma$ from ROD-mean. After application of all
cuts, 121 hours ($16\%$) of the total collected data survive. The
science channels display mean inter-channel cross-correlation
coefficients of $~1\%$ in the TOD, primarily caused by correlated
atmospheric fluctuations or HEMT gain fluctuations \citep{wol98}.
The QPC display similar correlations. Correlations between Stokes
parameters in the ROD average $10\%$, for the science channels
and $\sim 3\%$ for the QPC.
\begin{figure}[tb]
\centerline{\epsfxsize=9.8cm\epsffile{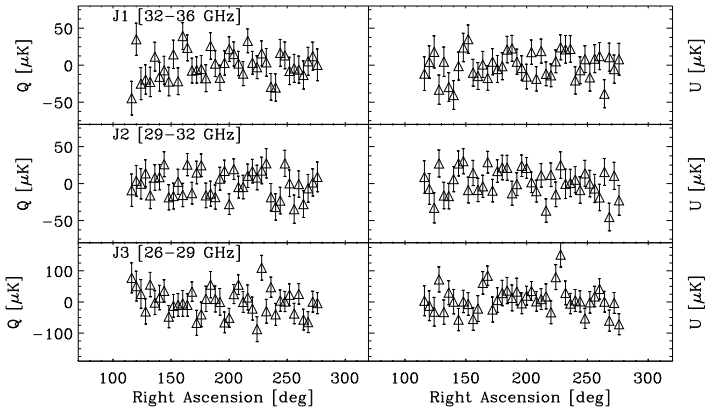}}
\caption{\label{f:maps}\footnotesize%
Maps of the Stokes parameters vs. R.A. for each frequency channel
in thermodynamic temperature. For display purposes, the maps are
binned into $4\arcdeg$ R.A. pixels with Galactic latitude $|b| >
25\arcdeg$, though the analysis uses $2\arcdeg$ pixels. The sky
coverage was determined by our scan strategy and data selection
criteria. The Stokes parameters are presented in accordance with
the IAU definition.} \vskip-0.5cm
\end{figure}
For each section of data that passes the data quality tests, the
time-ordered data are converted to a map and covariance matrix via
the minimum variance map-making procedure outlined in
\citet{teg97}. The 25 total sections range in length from 4-24
hours. An overall offset is removed from each section, and the
data covariance matrix is adjusted accordingly
\citep{teg97,bjk98}. All sections are combined into a final map
and covariance matrix for each channel.  Figure \ref{f:maps}
presents the maps of the Stokes parameters for all correlator
channels versus RA.

\section{Foregrounds}
The observations described were conducted over a wide range of
Galactic latitudes and therefore there is a potential for galactic
contamination, especially at low latitudes \citep{ben96,dav96}.
Galactic synchrotron emission can be up to $75\%$ polarized
\citep{ryb79}. No maps of polarized synchrotron emission exist at
$30$ GHz, and extrapolation of measurements at lower
frequencies, \eg\hspace{1 mm} \citet{bro76}, is not a reliable probe of
synchrotron polarization at $30$ GHz due to Faraday rotation.
Although the unpolarized intensity is not necessarily correlated
with the polarized intensity, we limit our
susceptibility to polarized synchrotron emission by only using
data corresponding to Galactic latitudes $|b|>25\arcdeg$,
\ie\hspace{1 mm} regions
of low unpolarized intensity.  For reference, extrapolation of the $408$ MHz
Haslam map assuming $10\%$ polarization over our observing region
with $b>25\arcdeg$ yields an RMS of only a couple of
$\mu$K.  This is consistent with our null result described below,
indicating that we see no evidence of foreground contamination.


\section{Likelihood analysis for Polarized CMB Fluctuations}
\label{s:like} 
\begin{figure}[tb]
\centerline{\epsfxsize=7.2cm\epsffile{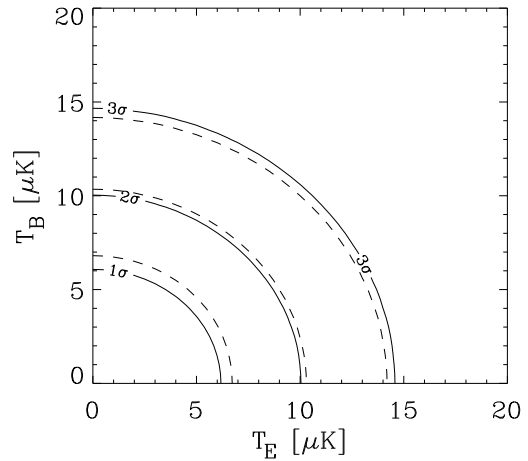}}
\caption{\label{f:likes}\footnotesize Normalized likelihood
contour plots in the $T_E-T_B$ plane. The countours enclose
are $68\%$, $95.4\%$, and $99.7\%$ of the total probability,
corresponding to $1, 2,$ and $3$
standard deviation intervals, as labeled. The solid lines
correspond to the joint likelihood for the combined three
frequency channels, and the dashed lines are the corresponding
null-channel (QPC) likelihood.} \vskip-0.5cm
\end{figure}

A Bayesian maximum likelihood analysis with a
uniform prior is used to test for polarized fluctuations. Since
both $Q$ and $U$ are measured simultaneously, we obtain joint
likelihoods for $E$ and $B$ in a model independent fashion
\citep{zal98}. The data from the three frequency channels are
combined into a single map for each Stokes parameter using the
minimum variance map-making procedure which includes all noise
correlations. We form a $2N_{\rm{pix}}$ element data vector
$\textbf{x}_i \equiv (Q_i, U_i)$ where $i = \{1 \ldots
N_{pix}\}$. The likelihood of the model given the data is
$\textsl{L}\propto \exp (-\x^T \C^{-1} \x/2)/\vert \C
\vert^{1/2}$. The covariance matrix
$\C_{ij}=\bf{S}_{ij}+\bf{N}_{ij}$, where $\bf{N}_{ij}=\lan \x^T_i
\x_j \ran$ is the noise covariance matrix which encodes
correlations for both Stokes parameters, at each pixel. The noise
covariance matrix also accounts for the offset removal required
to combine maps made over long timescales. $\bf{N}_{ij}$ is a
$2N_{\rm{pix}}\times 2N_{\rm{pix}}$ symmetric matrix, with
$N_{\rm{pix}}=84$ pixels of width $2\arcdeg$ for the $|b|>25\arcdeg$
Galactic latitude-cut data used for the CMB analysis.
$\bf{S}_{ij}$ is the signal covariance matrix constructed from a
simple flat band power model parametrized by $T_E, T_B$. To
place limits on the polarization we consider spectra for both $E$
and $B$ of the form $\ell(\ell+1) C^X_\ell/2\pi=T^2_X$ with $X
\in \{E,B\}$. The $Q$ and $U$ signal correlations are computed by
transforming the data coordinate basis to a local basis for each
pair of pixels. The local basis exploits the intrinsic symmetries
of the Stokes parameter correlation functions
\citep{zal98,kks97}. For each pixel pair the Stokes parameters in
the global basis are rotated such that the axis defining the
positive $Q$ direction of each pixel lies along the great-circle
connecting the pixels. The theory covariance matrix for all pairs
of Stokes parameters depends only on the sky coverage and the
underlying E and B-mode power spectra:
\vspace{2 mm}
\beq{cov} \bf{S}_{ij} =
\lan \x_i\x_j\ran =
\R(\alpha_{ij})\M(\rh_i\cdot\rh_j)\R(\alpha_{ji})^T. \eeq Here
$\M$ is the covariance matrix using a $(Q,U)$-convention where
the reference direction is the great circle connecting the two
points, and $\R(\alpha)$ are the rotation matrices defined by $
\R(\alpha)\equiv \left(\begin{tabular}{cc}
$\cos 2\alpha$     &$\sin 2\alpha$\\[2pt]
$-\sin 2\alpha$    &$\cos 2\alpha$\\[2pt]
\end{tabular}\right)$
 which rotate the Stokes parameters of the $\ith$ pixel into a
global reference frame where the reference directions are
meridians \citep{teg00}. $\bf{S}_{ij}$ is constructed from the
individual $2\times 2$ blocks of \eq{cov} by looping over all
pixel pairs, which requires
$N_{\textrm{pix}}(N_{\textrm{pix}}+1)/2$ operations. The
$\M$-matrix is given by \citep{zal98}
\beq{MdefEq}
\M(\rh_i\cdot\rh_j)\equiv \left(\begin{tabular}{cc}
$\expec{Q_i Q_j}$  &$0$\\[2pt]
$0$            &$\expec{U_i U_j}$\\[2pt]
\end{tabular}\right),
\eeq
where \beqa{CorrEq} \expec{Q_i Q_j}&\equiv&
  \sum_\l \left({2\l+1\over 4\pi}\right)
   B^2_\ell\left[F^{2}_{1,\l}(z) C_\l^E - F^{2}_{2,\l}(z) C_\l^B\right],\\
\expec{U_i U_j}&\equiv&
  \sum_\l \left({2\l+1\over 4\pi}\right)
   B^2_\ell \left[F^{2}_{1,\l}(z) C_\l^B - F^{2}_{2,\l}(z) C_\l^E\right],\label{CorrEq4}
\eeqa $z=\rh_i\cdot\rh_j$ is the cosine of the angle between the
two pixels under consideration,
$B_\ell=\exp[-\ell(\ell+1)\sigma^2_B/2]$, $\sigma_B$ is the beam
dispersion $=0.425\times\,$FWHM, and $F_{1},F_{2}$ are functions
of Legendre polynomials as defined in \citet{zal98} and \citet{teg00}. For
each point in the $T_E-T_B$ plane, equations 4 and 5 are
evaluated and substituted into equations 3 and 2 to compute
$\bf{S}_{ij}$. The joint likelihood for $\{T_E,T_B\}$ is shown in
Figure \ref{f:likes}. The $95\%$ confidence limits are obtained
by integrating the likelihood function to determine the region
which contains $95\%$ of the probability. The limits are
$T_E<10\mu$K and $T_B<10\mu$K at $95\%$ confidence for the
multipole range $2 < \ell < 20$. These limits do not include the
($\sim 10\%$) calibration uncertainty. Also shown in figure
\ref{f:likes} are the corresponding limits obtained from our null
channels; the QPC. Finally, since the B-mode power is expected to
be sub-dominant, even at large angular scales with no
reionization, we have also calculated the limits on $T_E$ with
$T_B\equiv 0$. Integrating the 1-D likelihood curve for $T_E$
with $T_B\equiv 0$ we find $T_E<8\mu$K at $95\%$ confidence.

\section{Discussion}
\polar probes the power spectra at large angular scales $2 < \ell
< 20$ where the signature of reionization and gravitational waves
are predicted to be most pronounced. While we find no evidence
for detection of either phenomenon, \polars limits are the most
restrictive upper limits at these angular scales, and represent
the first limits on E and B-modes from an experiment
simultaneously measuring both Q and U. In January 2001 \polar was
upgraded to measure CMB polarization at $20\arcmin$ scales. The
upgrade, called COsmic Microwave Polarization at Small Scales
({\scshape Compass}), uses the same polarimeter as {\polar} and
is an exciting complement to the results presented here. Assuming
similar per-pixel sensitivity to \polar and a generic E-mode
polarization spectrum, {\scshape Compass} has the potential to
detect polarization of the CMB and dramatically enhance the
scientific returns of this nascent field.




\acknowledgments We are grateful to Dick Bond, Robert
Brandenberger, Brendan Crill, Robert Crittenden, Khurram Farooqui,
Josh Gundersen, Wayne Hu, Kip Hyatt, Lloyd Knox, Arthur Kosowsky, Andrew Lange, Phil
Lubin, Melvin Phua, Alex Polnarev, Dan Swetz, David Wilkinson,
Grant Wilson, Ed Wollack, and Matias Zaldarriaga for many useful
conversations. BK and CO were supported by NASA GSRP
Fellowships.  \polars HEMT amplifiers were graciously provided by
John Carlstrom. This work has been supported by NSF grants AST
93-18727, AST 98-02851, and AST 00-71213, and NASA grant
NAG5-9194.




\end{document}